\documentclass[epjCONF]{svjour}
\usepackage{graphics}
\usepackage[varg]{txfonts} 
\usepackage[latin1]{inputenc}
\newcommand{\rrlyr} {RR~Lyr}
\newcommand{\kepl} {{\it Kepler}}

\newcommand{\pbla} {$P_B$}

\newcommand{\cd} {d$^{-1}$}
\newcommand{\bc} {$b_C$}
\newcommand{\rc} {$r_C$}
\newcommand{\gc} {$g_C$}

\session-title{Conference Title, to be filled}
\begin{document}
\title{The star RR Lyr and the Cepheid variables in the era of the space photometry revolution}
\author{Ennio Poretti\inst{1,2,3,4}
\and Jean-Fran\c{c}ois Le Borgne\inst{2,3,4} 
\and Alain Klotz\inst{2,3,4}}
\institute{INAF-Osservatorio Astronomico di Brera, Merate, Italy 
\and Universit\'e de Toulouse, UPS-OMP, IRAP, Toulouse, France 
\and Groupe Europ\'een d'Observations Stellaires (GEOS), Bailleau l'Ev\'eque, France
\and CNRS, IRAP, Toulouse, France 
}
\abstract{The long-term behaviours of the pulsation and Blazhko periods of RR~Lyr
are investigated by means of {\it Kepler} and ground-based observations. The difficulties
in detecting additional modes in the Cepheids monitored with CoRoT are discussed.
} 
\maketitle
%
\section{RR Lyr}
\label{sez1}
\begin{figure}
\begin{center}
\resizebox{1.0\columnwidth}{!}{
  \includegraphics{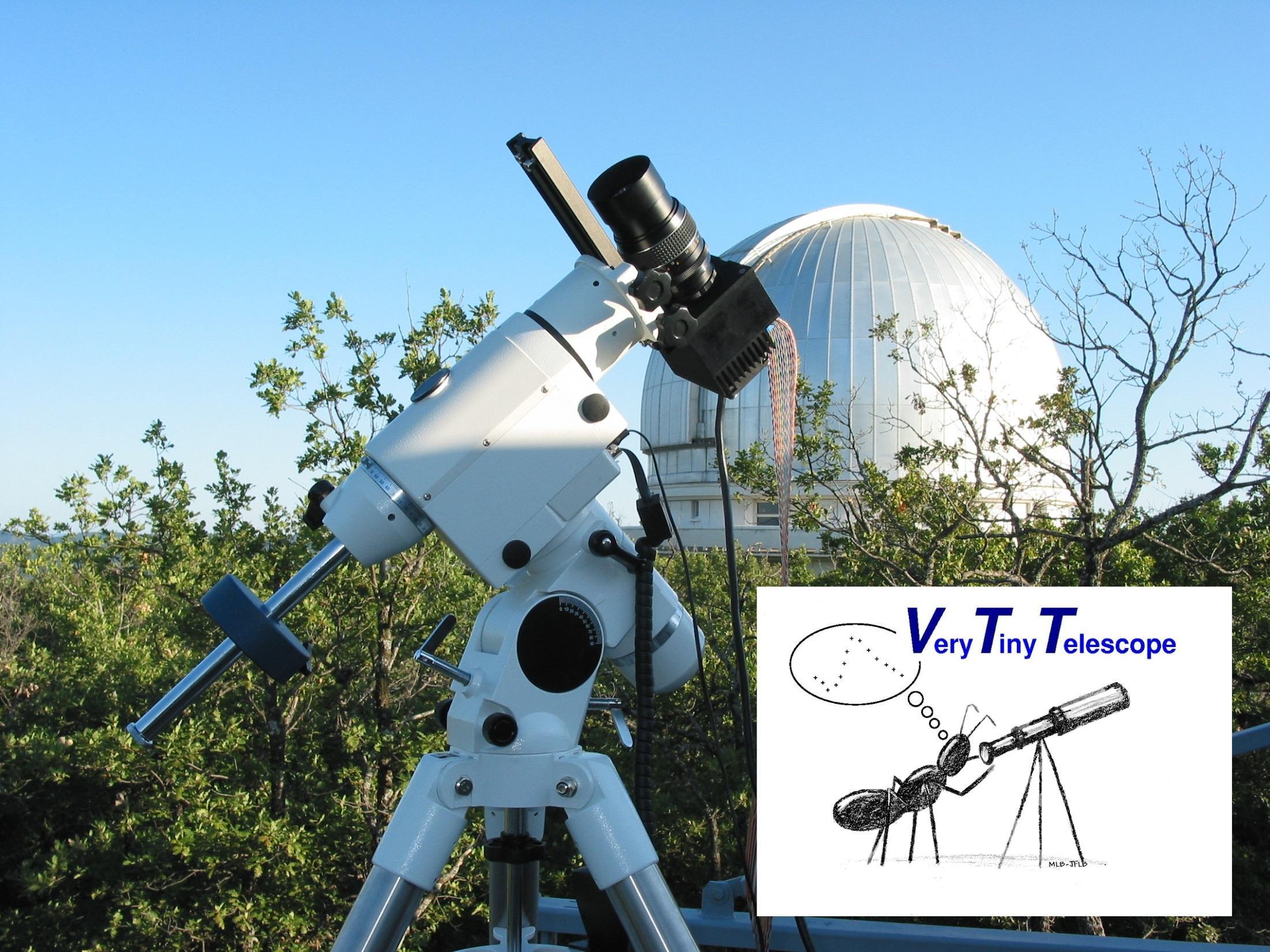}}
\caption{The Very Tiny Telescope (VTT). In the background the dome of the 1.93-m telescope
at the Observatoire de Haute-Provence.}
\label{vtt}       
\end{center}
\end{figure}

RR~Lyr$\equiv$KIC7198959 was included in
the field-of-view of the \kepl\, space telescope \cite{borucki} and high-precision, continuous
observations could be secured. Well before the launch of \kepl, we decided to devise small, 
autonomous, and transportable photometric instruments to make the ground-based survey of the 
modulations in amplitude and phase (Blazhko effect) of RR~Lyr as effective as possible.
The instruments are composed of a commercial equatorial mount
(Sky-Watcher HEQ5 Pro Goto), an AUDINE CCD camera (512x768 kaf400 chip) and a
photographic 135-mm focal, f/2.8 lens with a field of view of 2$^\circ$x3$^\circ$ (Fig.~\ref{vtt}).
We gave them the nickname VTTs for ``Very Tiny Telescopes". 

\begin{figure}
\begin{center}
\resizebox{1.0\columnwidth}{!}{
  \includegraphics{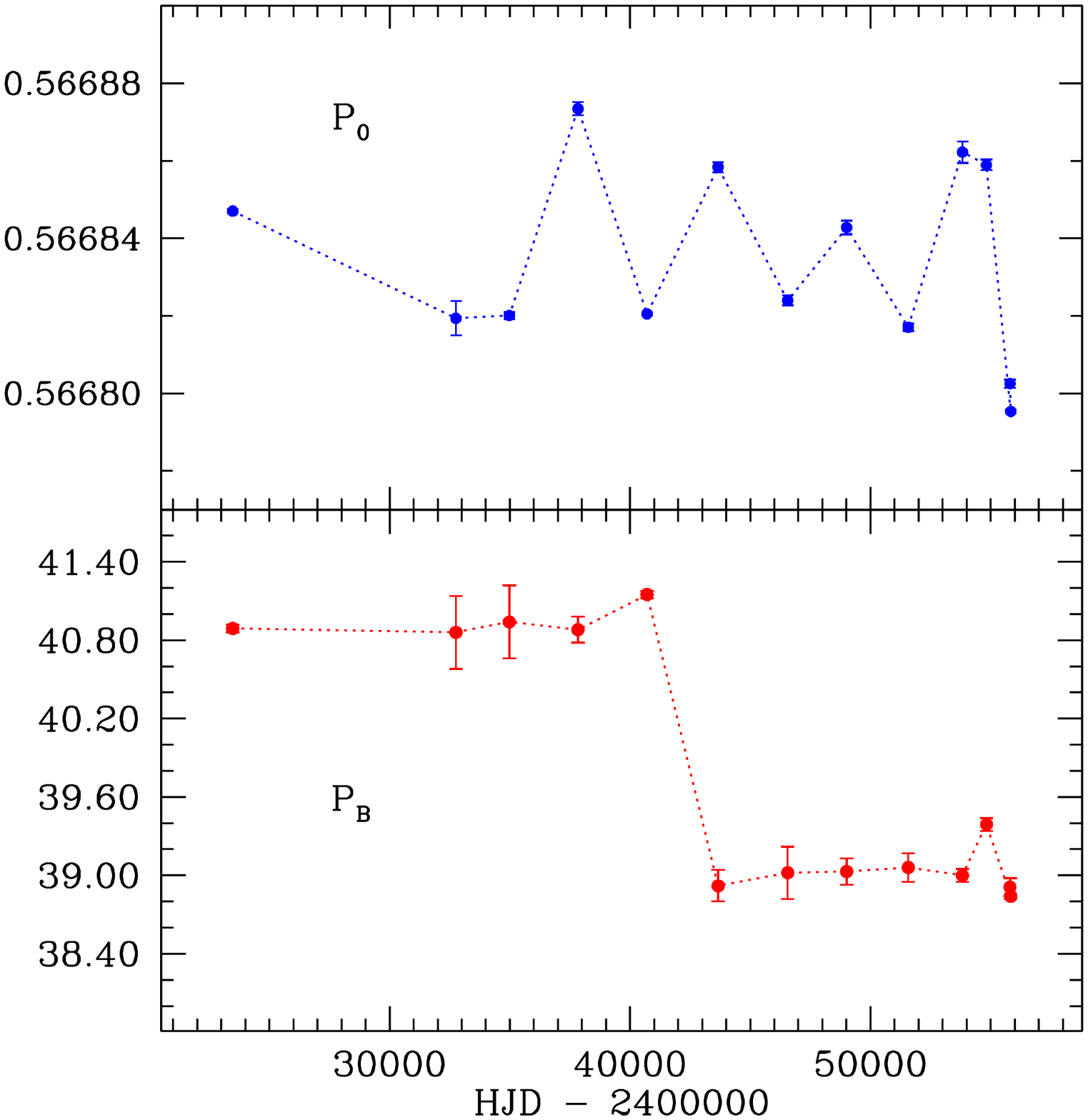} \includegraphics{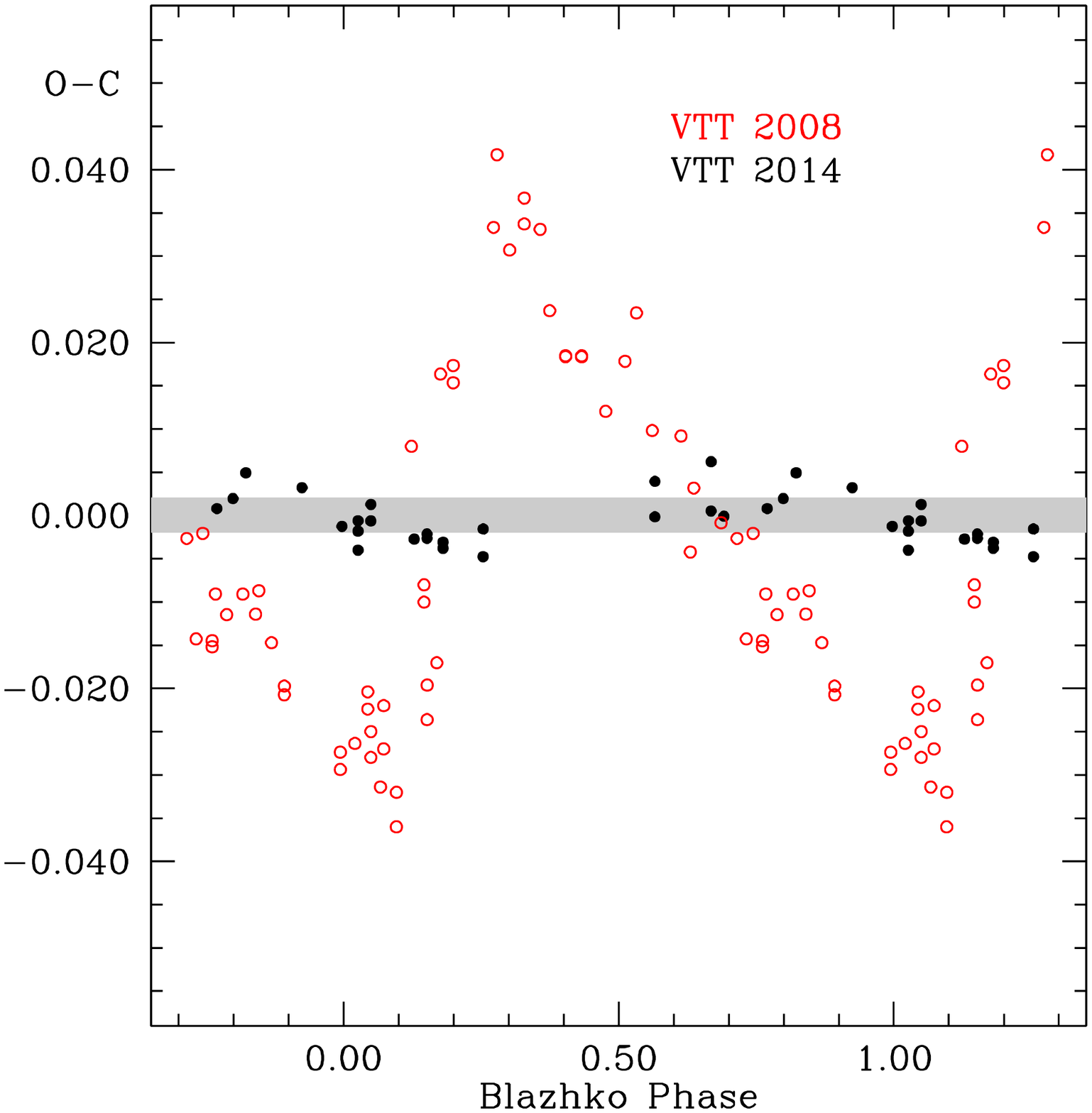}}
\caption{Measure units  of the ordinates are days. {\it Left panel}: decoupled variations of the pulsation period $P_0$ (top) 
and of the Blazhko period $P_B$ (bottom) from 1920 to 2013.
{\it Right panel}: the vanishing amplitude of the O-C values  from 2008 (0.062~d, red empty circles) to 2014 (0.002~d,
black filled circles). The gray band delimites the $\pm1\sigma$ interval around the mean of the O-C values
obtained in 2014. 
}
\label{deco}       
\end{center}
\end{figure}

The GEOS ({\it Groupe Europ\'een d'Observations Stellaires}) RR~Lyr database
\cite{pervar}
is continuously updated by inserting all the published maxima of \rrlyr\, variables.
\rrlyr\, itself is among them 
\footnote{http://rr-lyr.irap.omp.eu/dbrr/dbrr-V1.0\_08.php?RR\%20Lyr)}.
Therefore, we could analyse \cite{vttpaper} all the observed maxima since 1899 
in order to have a clear picture of the behaviours of the pulsation period $P_0$ and of the 
Blazhko effect, noticed since 1916.
We could identify
two states defined as pulsation over a ``long" primary period 
($P_0>$0.56684~d) and over a ``short" one ($P_0<$0.56682~d).
These states alternate with intervals of 13-16~yr, and are well defined since  1943. 
We also provided homogeneous determinations of the Blazhko period $P_B$ in several time intervals and
we studied how it changed while the two states alternated.
We could clearly establish how the Blazhko period  had just one sudden decrease from 40.8~d to 
39.0~d in 1975.
The variations of the pulsation  and  Blazhko periods are completely decoupled (Fig.~\ref{deco}).

Moreover, the combination of \kepl\, and VTT data recorded the vanishing of the Blazhko
effect. 
The space telescope  continuously monitored the monotonic long-term decrease, proving that
small-scale modulations, lasting from 2 to 4 \pbla, are also visible  in the O-C 
({\it observed} minus {\it calculated} times of maximum brightness) values.
The VTT data are now securing the \rrlyr\, monitoring after that \kepl\, had to be pointed in another direction.
In particular, Fig.~\ref{deco} shows how the large sine-shaped curve of the O-C values observed in 2008 
turned out to be an almost straight line in 2014.
The Blazhko nature of the light variability of \rrlyr\, is hard to detect by looking only to the maxima
collected in the year 2014.





\section{Cepheids}
Though nonradial modes were already detected in classical pulsators from ground-based surveys, it has
been the space revolution that provided us clear cases of the excitation of nonradial modes in
high-amplitude $\delta$ Sct stars and RR Lyr variables. In this revolutionary context, Cepheids are
playing the role of the conservative party, since additional modes are very rare in their light
curves. Low amplitude frequencies were found in OGLE-II stars \cite{pavel}, more
specifically secondary peaks in proximity to the dominant first overtone radial mode. The stars could
be both single-periodic first-overtone Cepheids and double-mode ones. Intriguing cases are also those of 
Cepheids showing one secondary mode with frequency much above that of the radial mode. The situation
did not change after the \kepl\, pointing in the Lyra-Cygnus direction: there was only one Cepheid in the
field-of-view, i.e., V1154~Cyg. The detailed
analysis of the data acquired in the first 600~days (i.e., about
120 cycles) of the mission detected significant cycle-to-cycle fluctuations \cite{v1154}.
A very slight correlation was also found between the Fourier parameters and
the O-C values, suggesting that the O-C 
variations (up to 30~min) might be due to
instabilities in the light curve shape. This correlation was strengthened by the analysis of the
\kepl\, measurements up to Q17 \cite{kanev}.

Recently we re-investigated all the variables observed by CoRoT \cite{esa3} in the exoplanetary field. 
We were able to detect eight Cepheids. The most important case was that of CoRoT 0223989566,
that resulted to be an unusual  triple-mode Cepheid located very probably in the ``outer
arm" of the Milky Way \cite{triple}. Two other Cepheids were observed in Long Runs and therefore
we could perform a detailed frequency analysis to investigate the excitation of additional modes.
The case of CoRoT 0102618121 was particularly relevant in this context. After subtracting the main oscillation,
the residuals obtained both in the LRa01 (131~d long) and in the LRa06 (76~d) data revealed a possible peak
at 0.488~\cd. However, when computing the residuals in another way, i.e., cycle-by-cycle, this peak
disappeared from the power spectra of both runs. Indeed, the light curve obtained from plotting the cycle-by-cycle residuals
is completely flat and very different from that from the fit of the entire dataset. 
We also notice that  we did not detect any linear combination between the main pulsation frequency and the 0.488~\cd\, one:
this fact does not support the identification of the latter term as an  additional mode.
Our working hypothesis is that
the light curve is affected by very small jumps and drifts, leaving some traces of the large amplitude, main oscillation
after removing a solution with fixed amplitudes, phases and frequencies from the data. 
\begin{figure}
\begin{center}
\resizebox{0.75\columnwidth}{!}{
  \includegraphics{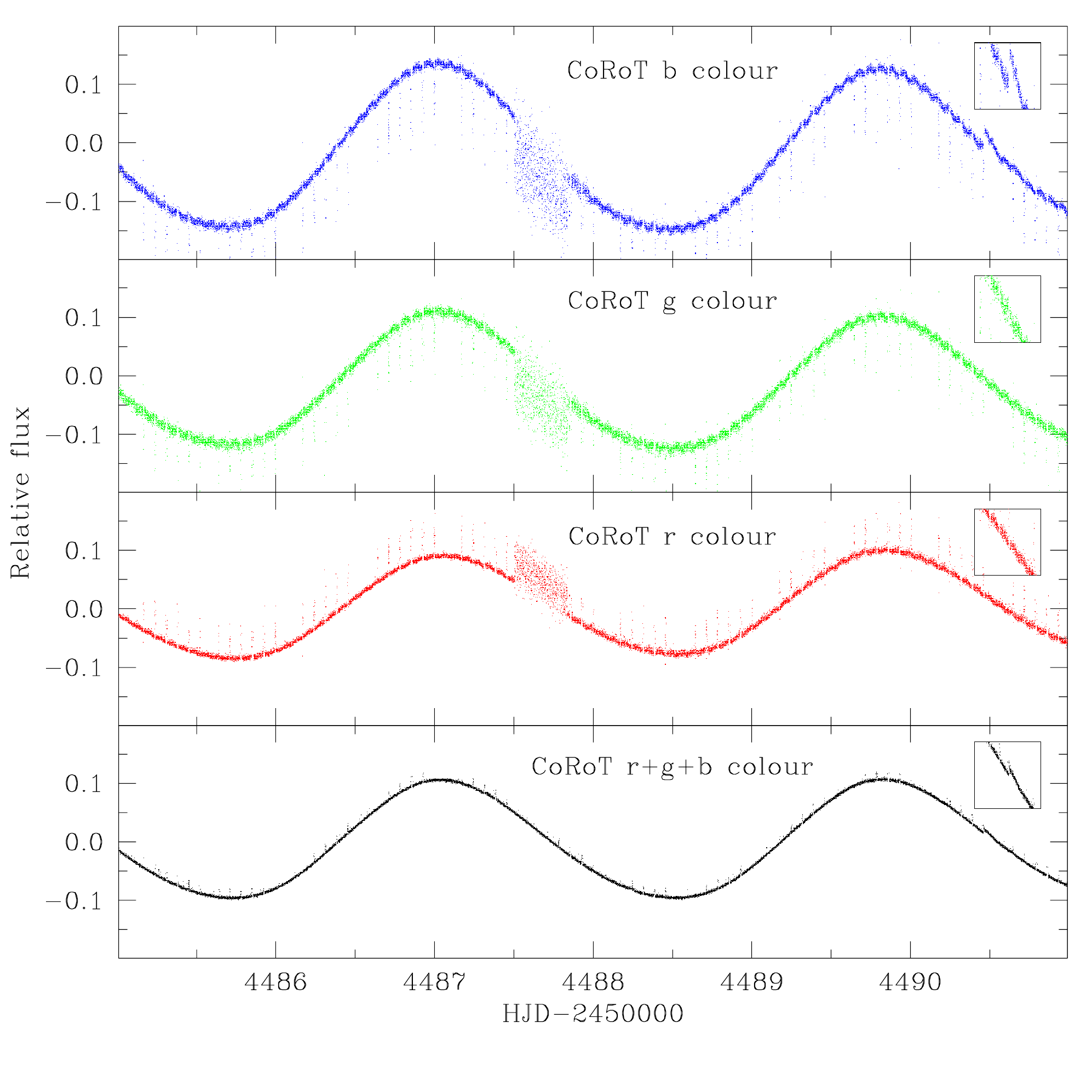} }
\caption{
The CoRoT light curves in \rc, \gc, \bc\, and white colours reflect in different ways the effect
of the satellite jitter around JD~2454487.5. The jump around JD~2454490.5 (inserted boxes)
was observed in the \bc-light, but not in the \rc\, and \gc\, ones.
}
\label{ciclo}       
\end{center}
\end{figure}

As an example of these subtle instrumental effects, let us consider the light curves obtained by CoRoT in the \rc, \gc, and \bc\, colours.
The most relevant perturbation occurred around JD~2454487.5 (Fig.~\ref{ciclo}): a sort of
blurring of the light curve that lasted for about 0.5~d.
We can suppose that the star's image jumped all around  due to a sudden increase of the satellite jitter.
As a result, the flux subdivision changed: the counts decreased in the \bc\, and \gc\, curves,
while increased  in the \rc\, one. However, the total counts  did not change significantly  and the white
light curve perfectly describes the star's variability. A different event 
occurred one cycle later. The \bc\, curve shows a sudden jump, followed
by a decline to the expected flux level.  This event was not recorded in \gc\, and \rc\,
colours (zoom-in in the inserted boxes).  Probably a cosmic ray  hit a pixel in the \bc\, submask. 
Therefore, the curve in white light shows
the event on a reduced scale with respect to the \bc\, one. These abrupt changes are hard
to be  noticed by eye and the automatic removal is quite dangerous. Indeed,
rapid variations are what we are willing to put in evidence with the goal to  detect possible planetary transits. 
On the contrary, the cycle-to-cycle
fits minimize such effects leaving a light curve cleaner than that of the fit of the entire dataset. 
We cannot rule out that we are
actually observing small cycle-to-cycle variations due to some physical mechanism not strictly periodic, as granulation
or stellar activity.  Further analyses are in progress \cite{irap}.
%

\begin{acknowledgement}
EP acknowledges Observatoire Midi-Pyr\'en\'ees for the two-month grant
allocated between 2014 May and July, allowing him to  spend a very fruitful  stage at the
{\it Institut de Recherche en Astrophysique et Plan\'etologie} in Toulouse, France.
\end{acknowledgement}

\end{document}